\documentclass{PoS}
\usepackage{mathrsfs}
\title{Second Order Phase Transition in Anisotropic Lattice Gauge Theories with Extra Dimensions}

\ShortTitle{Second Order Phase Transition in Anisotropic Lattice Gauge Theories\ldots}

\author{\speaker{Stam Nicolis}%
        \\
        CNRS-Laboartoire de Math\'ematiques et Physique Th\'eorique (UMR
        6083)\\
        F\'ed\'eration de Recherche ``Denis Poisson'' (FR 2964)\\
        Universit\'e ``Fran\c{c}ois Rabelais'' de Tours\\
        Parc Grandmont, 37200 Tours, France\\
        E-mail: \email{Stam.Nicolis@lmpt.univ-tours.fr}}

\abstract{
Field theories with extra dimensions live in a limbo. While their classical
solutions have been the subject of considerable study, their quantum aspects
are difficult to control. A special class of such theories are 
anisotropic gauge theories. The anisotropy was originally  introduced to 
localize chiral fermions.  Their continuum limit is of practical interest and
 it will be shown that the anisotropy of the gauge couplings plays a 
crucial role in opening the phase diagram of the theory to a new phase, that is separated from the others by a second order phase transition. The mechanism behind this is   generic for a certain class of models, that can be studied with lattice techniques. This leads to new perspectives for the study of quantum effects of extra dimensions.
}

\FullConference{The XXVIII International Symposium on Lattice Filed Theory\\
		 June 14-19,2010\\
		 Villasimius, Sardinia Italy}

\begin{document}
Extra dimensions\footnote{More than the three spatial dimensions we typically
  perceive. We stick to one time dimension.} lead a troubled existence in
field theory, since quantum field theories in more than four dimensions
are plagued by untameable ultraviolet divergences, rendering any
calculation, beyond the solution of the equations of motion, 
 sensitve to the details of the regularization used. The lattice
 regularization has the advantage that it can be used to probe perturbative as
 well as non-perturbative aspects and thus provide hints about this
 sensitivity. We will use it to study a particular class of theories,
 anisotropic gauge theories~\cite{fu_nielsen} (with compact gauge group). We
 want to see what effect the anisotropy has on the order of the transitions
 between the various phases (so we specialize to the case of compact
 $U(1)$). While there has been a fair amount of numerical
 work~\cite{berman_rabinovici,HKAN,hulsebos,farakos_vrentzos}, the reason {\em why} 
 a second order phase transition should, indeed, appear, has not been really
 spelled out. So it's useful to see how this could happen in a concrete
 example--as well as what could prevent its appearence.
 In addition, the method used has a wider applicability and deserves 
 being recalled. 

\begin{figure}[thp]
\includegraphics[scale=0.8]{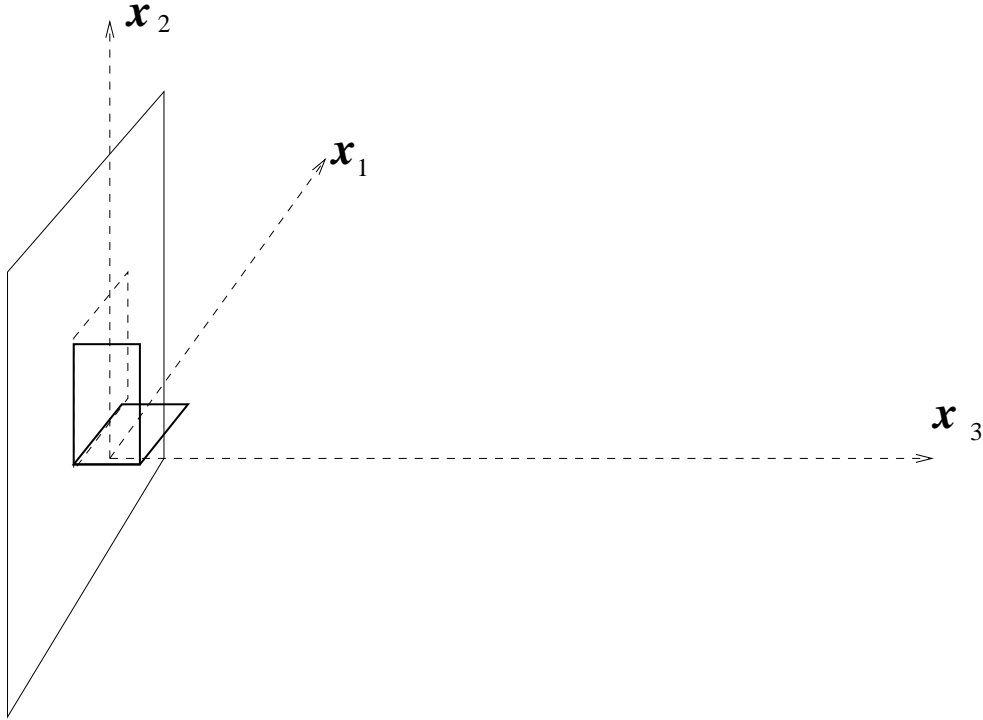}
\caption[]{An example in $D=d_\parallel+d_\perp=3$, with $d_\parallel=2$ and
  $d_\perp=1$. The plaquettes in the $(x_2,x_3)$ and $(x_3,x_1)$ planes 
enter with 
 coefficient $\beta'$ in the action; the plaquettes in  the $(x_1,x_2)$ plane enter with coefficient $\beta$.}
\label{anisot_space}
\end{figure}

The action is in Wilson form
\begin{equation}
\label{anisot_action}
S = \beta\sum_n\sum_{1\leq\mu<\nu\leq d_\parallel}
\left(1 - \mathrm{Re}\left[U_{\mu\nu}\right]\right) + 
\beta'\sum_n\sum_{d_\parallel+1\leq\mu<\nu\leq d_\parallel+d_\perp}
\left(1 - \mathrm{Re}\left[U_{\mu\nu}\right]\right) 
\end{equation}
corresponding to the situation illustrated in fig.~\ref{anisot_space} for
$d_\parallel=2$ and $d_\perp=1$
We shall use a technique developed for implementing the mean field
approximation to systems with local symmetries 
 to obtain the phase diagram, namely
a trick introduced in ref.~\cite{brezin_drouffe} (we use it in the form
presented in~\cite{fu_nielsen,flyvbjerg}).

We insert in the partition function,
\begin{equation}
\label{partition_fun1}
Z[J] = \int [{\mathscr D} U] e^{-S[U]+\sum_n \mathrm{Re}(J_\mu(n) U_\mu(n))}  
\end{equation}
the expression
\begin{equation}
\label{deltafunction}
1 = \int  \left[\prod_\mathrm{links}\int d\mathrm{Re}(V_l)d\mathrm{Im}(V_l)
\delta(\mathrm{Re}(V_l)-\mathrm{Re}(U_\mu(n)))   
\delta(\mathrm{Im}(V_l)-\mathrm{Im}(U_\mu(n)))   
\right]
\end{equation}
to decouple the gauge links
\begin{equation}
\begin{array}{l}
\displaystyle
Z[J] = 
\int [{\mathscr D} U] 
\left[\prod_\mathrm{links}\int d\mathrm{Re}(V_l)d\mathrm{Im}(V_l)
\delta(\mathrm{Re}(V_l)-\mathrm{Re}(U_\mu(n)))   
\delta(\mathrm{Im}(V_l)-\mathrm{Im}(U_\mu(n)))   
\right]
e^{-S[U]+\sum_n \mathrm{Re}(J_\mu(n) U_\mu(n))}=\\
\displaystyle
\int [{\mathscr D} U] 
\left[\prod_\mathrm{links}\int d\mathrm{Re}(V_l)d\mathrm{Im}(V_l)
\frac{d\alpha_l^R}{2\pi}\frac{d\alpha_l^I}{2\pi}
e^{\mathrm{i}\sum_l\mathrm{i}\alpha_l^R(-\mathrm{Re}(V_l)+\mathrm{Re}(U_\mu(n)))}
e^{\mathrm{i}\sum_l\mathrm{i}\alpha_l^I(-\mathrm{Im}(V_l)+\mathrm{Im}(U_\mu(n)))}
\right]
e^{-S[U]+\sum_n \mathrm{Re}(J_\mu(n) U_\mu(n))} = \\
\displaystyle
\int
\left[\prod_\mathrm{links}\int d\mathrm{Re}(V_l)d\mathrm{Im}(V_l)
\frac{d\alpha_l^R}{2\pi}\frac{d\alpha_l^I}{2\pi}\right]
e^{-S[\mathrm{Re}(V_l),\mathrm{Im}(V_l)]+\sum_l(\mathrm{Re}(J_l)\mathrm{Re}(V_l)-\mathrm{Im}(J_l)\mathrm{Im}(V_l))-\mathrm{i}\sum_l\alpha_l^R\mathrm{Re}(V_l)-\mathrm{i}\sum_l\alpha_l^I\mathrm{Im}(V_l)
+\sum_l w(\alpha_l^R,\alpha_l^I)}
\end{array}
\end{equation}
where $w(\alpha_l^R,\alpha_l^I)$ contains the information about the gauge group,
\begin{equation}
\label{gauge_int}
e^{w(\alpha_l^R,\alpha_l^I)}\equiv 
\int{\mathscr D} U e^{\mathrm{i}(\alpha_l^R\mathrm{Re}(U_\mu)+\alpha_l^I\mathrm{Im}(U_\mu))}
\end{equation}
So far we have an exact transcription: we have traded the {\em constrained}
variables, $U_\mu(n)$ (that must satisfy $[\mathrm{Re}(U_\mu(n))]^2 +
[\mathrm{Im}(U_\mu(n))]^2=1$), for the {\em unconstrained} variables, $\alpha_l^R,
\alpha_l^I, \mathrm{Re}(V_l),\mathrm{Im}(V_l)$.  It is, indeed, the existence
of the constraint that leads to a non-trivial dependence on the coupling
constant(s) of the effective action thus obtained, already at the
``classical'' level. 

The effective action seems to have acquired terms that are complex--however
the way they enter allows us to perform a ``Wick rotation'',
$\mathrm{i}\alpha_l^R\equiv
\widehat{\alpha}_l^R,\mathrm{i}\alpha_l^I\equiv\widehat{\alpha}_l^I$ and
obtain an action that is manifestly real:
\begin{equation}
\label{unconstrained}
S_\mathrm{eff}(\widehat{\alpha}_l^R,\widehat{\alpha}_l^I,\mathrm{Re}(V_l),\mathrm{Im}(V_l))
= S[\mathrm{Re}(V_l),\mathrm{Im}(V_l)] +
\sum_l\left(\widehat{\alpha}_l^R\mathrm{Re}(V_l) + \widehat{\alpha}_l^I\mathrm{Im}(V_l)\right) -
\sum_l w(\widehat{\alpha}_l^R,\widehat{\alpha}_l^I)
\end{equation}
We can, in fact, use this action for Monte Carlo simulations--but, also, for
analytical computations, that are much easier to perform, since we have solved
the constraints~\cite{brezin_drouffe,flyvbjerg,fl_lautrup_zuber}.

We now specialize eq.~\ref{unconstrained} to  the case of the action in eq.~(\ref{anisot_action}) and look for extrema
that are uniform along the $d_\parallel$--dimensional respectively along the 
$d_\perp$ extra dimensions: $V_l\equiv v$, for links that belong in the
$d_\parallel$--dimensional subspaces and $V_l\equiv v'$ for links that ``point
out'' along the $d_\perp$ extra dimensions. Similarly $\widehat{\alpha}_l
\equiv \widehat{\alpha}$ within the $d_\parallel$ dimensional subspaces and 
$\widehat{\alpha}_l\equiv\widehat{\alpha}'$ along the $d_\perp$ extra
dimensions. A plaquette that lies in the $d_\parallel$--dimensional subspace
makes the following contribution to the effective action
\begin{equation}
\label{dparallelplaq}
\left.\mathrm{Re}[U_{\mu\nu}(n)]\right|_{1\leq\mu<\nu\leq d_\parallel}
 = \mathrm{Re}[(v^R +
  \mathrm{i}v^I)(v^R+\mathrm{i}v^I)(v^R-\mathrm{i} v^I)(v^R-\mathrm{i} v^I)]=
([v^R]^2 + [v^I]^2)^2
\end{equation}
Similarly, a plaquette that lies in the $d_\perp$--dimensional subspace
contributes the expression
\begin{equation}
\label{dperpplaq}
\left.\mathrm{Re}[U_{\mu\nu}(n)]\right|_{d_parallel+1\leq\mu<\nu\leq
  d_\parallel+d_\perp} = ([v'^R]^2 + [v'^I]^2)^2
\end{equation}
A plaquette that ``spans'' the subspace between two $d_\parallel$--dimensional
subspaces contributes
\begin{equation}
\label{mixedplaq}
\left.\mathrm{Re}[U_{\mu\nu}(n)]\right|_{1\leq\mu\leq d_\parallel<\nu\leq
  d_\parallel+d_\perp} = \mathrm{Re}( 
(v^R+\mathrm{i}
v^I)(v'^R+\mathrm{i}v'^I)(v^R-\mathrm{i}v^I)(v'^R-\mathrm{i}v'^I) )) = 
([v^R]^2+[v^I]^2)([v'^R]^2 + [v'^I]^2)
\end{equation}
Simple counting allows us to write down the expression for the effective
action for such uniform configurations:
\begin{equation}
\begin{array}{l}
\displaystyle
S_\mathrm{eff}[v^R,v^I,v'^R,v'^I,\widehat{\alpha}^R,\widehat{\alpha}^I,
\widehat{\alpha'}^R,\widehat{\alpha'}^I] = 
\beta\frac{d_\parallel(d_\parallel-1)}{2}\left(1-([v^R]^2+[v^I]^2)^2\right) + 
\beta'\frac{d_\perp(d_\perp-1)}{2}\left(1-([v'^R]^2+[v'^I]^2)^2\right) + \\
\displaystyle
\beta'd_\perp d_\parallel\left(1-([v^R]^2+[v^I]^2)([v'^R]^2 +
       [v'^I]^2)\right)+\\
\displaystyle 
d_\parallel
\left(\widehat{\alpha}^R v^R + \widehat{\alpha}^I v^I -
w(\widehat{\alpha}^R,\widehat{\alpha}^I)\right) +
d_\perp \left(\widehat{\alpha'}^R v'^R + \widehat{\alpha'}^I v'^I -
w(\widehat{\alpha'}^R,\widehat{\alpha'}^I)\right)
\end{array}
\end{equation}
For the case of compact $U(1)$ the gauge group integral is given in terms of
elementary functions:
\begin{equation}
\label{gauge_int_U1}
e^{w(\widehat{\alpha}^R,\widehat{\alpha}^I)} = 
\int_{-\pi}^{\pi}\frac{d\theta}{2\pi} e^{\widehat{\alpha}^R\cos\theta +
  \widehat{\alpha}^I\sin\theta} = \int_{-\pi}^{\pi}\frac{d\theta}{2\pi}
e^{\sqrt{[\widehat{\alpha}^R]^2 +
    [\widehat{\alpha}^I]^2}\cos(\theta-\phi_{\widehat{\alpha}})}  \equiv 
I_0\left(\sqrt{[\widehat{\alpha}^R]^2 +
    [\widehat{\alpha}^I]^2}\right)
\end{equation}
where $I_0(\cdot)$ is the modified Bessel function. 

We notice that the group integral depends only on the length of the
``vector(s)''$(\widehat{\alpha}^R,\widehat{\alpha}^I)$--and that the plaquette
terms in the effective action depend only on the length of the ``vector(s)''
$(v^R,v^I)$. The two vectors are coupled only through their ``scalar product'',
$\widehat{\alpha}^R v^R + \widehat{\alpha}^Iv^I$, which depends on their
lengths and their
{\em relative} orientation. This means that we can {\em choose} a convenient
coordinate system in this  space and, as long as the
corresponding symmetry isn't spontaneously broken, we can simplify the
calculations considerably. 
We thus choose the orientations so that $v^I = 0$, $v'^I = 0$,
$\widehat{\alpha}^I=0$, $\widehat{\alpha'}^I = 0$. Indeed we easily check that 
this choice is a solution of the equations for the extrema of the effective
action. In a sense this amounts to ``choosing a gauge'' in this theory. To
simplify notation we henceforth set $v^R\equiv v$, $v'^R\equiv v'$,
$\widehat{\alpha}^R\equiv \widehat{\alpha}$, $\widehat{\alpha'}^R\equiv
\widehat{\alpha'}$. 

In this ``gauge'', therefore, the action takes the form
\begin{equation}
\label{gauge_fixed_action}
\begin{array}{l}
\displaystyle
S_\mathrm{eff}[v,v',\widehat{\alpha},\widehat{\alpha'}] = 
\beta\frac{d_\parallel(d_\parallel-1)}{2}\left(1-v^4\right) + 
\beta'\frac{d_\perp(d_\perp-1)}{2}\left(1-v'^4\right) + 
\beta'd_\parallel d_\perp\left(1-v^2 v'^2\right) + \\
\hskip3truecm
\displaystyle 
d_\parallel(\widehat{\alpha} v - w(\alpha)) +
d_\perp(\widehat{\alpha'}v'-w(\alpha'))
\end{array}
\end{equation}
Compactness of the gauge group implies that $w(0)=1$ and $\infty>w''(0)>0$. In
addition, $w'(0)=0$. These features may be seen to hold for compact
$U(1)$--but they hold for {\em any} compact group.

The extrema of the effective action are solutions of the equations
\begin{equation}
\label{eom}
\begin{array}{l}
\displaystyle v = dw(\widehat{\alpha})/d\widehat{\alpha}\\
\displaystyle v' = dw(\widehat{\alpha'})/d\widehat{\alpha'}\\
\displaystyle \widehat{\alpha} = 2\beta d_\parallel(d_\parallel-1)v^3+2\beta'd_\parallel
d_\perp v v'^2\\
\displaystyle \widehat{\alpha'} = 2\beta'd_\perp(d_\perp-1)v'^3+2\beta'd_\parallel v^2 v'
\end{array}
\end{equation}
These equations always possess the solution
$(\widehat{\alpha},\widehat{\alpha'},v,v')=(0,0,0,0)$ that corresponds to the
confining phase--the string tension is infinite. However they also have
non-zero solutions, that depend on the values of the couplings $\beta$ and
$\beta'$. 
 The reason this is possible is 
that uniform configurations are only
invariant under global (constant) gauge transformations--and Elitzur's theorem~\cite{elitzur} 
holds only if local transformations are possible. Thus it is  {\em not} a
contradiction of Elitzur's theorem but rather a consequence of the fact that
the assumption behind it does not hold for the configuration under study. 

We thus find a solution with $(\widehat{\alpha},\widehat{\alpha'},v,v')\neq
(0,0,0,0)$, which corresponds to a $d_\parallel+d_\perp$--dimensional 
Coulomb phase (since Wilson loops with perimeter $L=L_1+L_2$ behave as $v^L$,
$v'^L$ or $v^{L_1}v'^{L_2}$). 

However there also exists a solution with 
$\widehat{\alpha}\neq 0,v\neq 0,\widehat{\alpha'}=0,v'=0$. In this phase (named
the ``layered phase'' in ref.~\cite{fu_nielsen}) the Wilson loops show
perimeter behavior within a $d_\parallel$--dimensional subspace (since $v\neq
0$) and show confinement along the $d_\perp$ directions, since $v'=0$. There
isn't any ``bulk'' at all: the $d_\parallel+d_\perp$--dimensional space has
become a stack of $d_\parallel$--dimensional layers. Since the string tension
is infinite the layers are infinitely thin and the theory on them is
local. Corrections to the mean field approximation will make this string
tension finite--the layers will acquire a thickness, inversely proportional to
the (square root of the) string tension and the theory will display {\em
  non-local} features, if probed at such length scales. For this to be
consistent this string tension should be much larger than the tension of the
fundamental string. 

 In all cases considered here the
boundary conditions are assumed to be periodic, but all dimensions are assumed
to become infinite in the continuum limit.

It is interesting to try and see whether the transition from one phase to
another can become continuous. Indeed the mean field approximation to lattice
gauge theories typically predicts first order (discontinuous transitions). The
reason can be understood from the expression of the action: the plaquette
terms, in the isotropic case, are quartic in the link variables. The only
terms that can contribute to quadratic order are the ``constraint'' terms, 
$\widehat{\alpha} v - w(\widehat{\alpha})$. If we replace 
$v=dw(\widehat{\alpha})/d\alpha$ and expand to quadratic order, 
around $\alpha=0$, we find that this point corresponds to a minimum of the 
effective action,that can never become a maximum. Therefore, 
if another minimum appears for $\alpha\neq 0$, the transition is, necessarily, 
of first order. 
Such a minimum, corresponding to a Coulomb phase, is only credible  for a 
theory with a $U(1)$ factor: the
putative Coulomb phase turns out to be an artefact of the mean field
appoximation~\cite{fl_lautrup_zuber} for Yang--Mills theories with a simple
Lie group and Monte Carlo simulations indicate that they 
are always confining at strong coupling and asymptotically free at
 weak couling~\cite{creutz}. 

In the case under study here, however, there is a term in the action that {\em
  can} destabilize the confining phase in a way consistent with a continuous
transition: the term
\begin{equation}
\label{destabilization}
S_\mathrm{eff}^{\mathrm{mixed}} = \beta'd_\parallel d_\perp (1 - v^2 v'^2)
\end{equation}
is quadratic in the link variables, due to the anisotropy. And these variables
enter with a sign that allows them to destabilize the confining phase along
the $d_\perp$ directions. To see this we expand the effective action around
the solution $(\widehat{\alpha},\widehat{\alpha'}=0)$, which exists for
$\beta$ large enough and $\beta'$ small enough, within the subspace where 
$v=dw(\widehat{\alpha})/d\widehat{\alpha}$ and
$v'=dw(\widehat{\alpha'})/d\widehat{\alpha'}$. So we consider $\alpha'$ small
enough that we may expand around $\widehat{\alpha'}=0$
 to quadratic order--but we retain the exact dependence on $\alpha$. We find 
\begin{equation}
\label{destabilize}
S_\mathrm{eff}[v,v',\widehat{\alpha},\widehat{\alpha'}]\approx 
S_\mathrm{eff}[v,0,\widehat{\alpha},0] +
\widehat{\alpha}'^2 w''(0)d_\perp\left[-\beta'd_\parallel v^2(\widehat{\alpha})w''(0) + 
\frac{1}{2}\right] 
\end{equation}
This expression depends on $\beta$ implicitly, since
$\widehat{\alpha}=\widehat{\alpha}(\beta)$. If $v(\widehat{\alpha})\neq
0$--the system is in the Coulomb phase within a $d_\parallel$--dimensional
subspace--there is a line, 
\begin{equation}
\label{critical_line}
\beta'_\mathrm{crit}(\beta)
 = \frac{1}{2 d_\parallel v^2(\widehat{\alpha})w''(0)}
\end{equation} 
such that, for $\beta'<\beta'_\mathrm{crit}$ the system is in the layered
phase and for $\beta'>\beta'_\mathrm{crit}$ it is in the
$d_\parallel+d_\perp$--dimensional Coulomb phase through a continuous
transition. 
For $U(1)$, in particular, $w''(0)=1/2$ and $v(\widehat{\alpha})$
is a bounded faunction of $\widehat{\alpha}(\beta)$, that tends to 1 as
$\widehat{\alpha}(\beta)\to\infty$. In that limit, which is relevant as
$\beta\to\infty$, we obtain that $\beta'_\mathrm{crit}\to 1/d_\parallel$, a
result that is compatible with the mean field approximation, which may be
considered an expansion in $1/d_\parallel$ (and was found in another way in
ref.~\cite{fu_nielsen}).  
This has further interesting consequences since, many years ago,
Peskin~\cite{peskin} noted that at a second order phase transition point the
static quark--anti-quark potential, derived from the Wilson loop, would
display $1/R$ behavior independently of the dimensionality. 
To date an example of such a system was not
available. Anisotropic lattice gauge theories with a $U(1)$ factor could provide
such an example and it will be interesting to explore its consequences further
through Monte Carlo simulation.

In conclusion we have identified the mechanism that is behind the transition
from the layered phase to the bulk Coulomb phase for anisotropic lattice gauge
theories, whose symmetry group contains a $U(1)$ factor. Monte Carlo
simulations to check its validity beyond the mean field approximation are
certainly called for and the theory in the continuum limit, 
especially in the presence of matter, remains to be constructed. Theories with 
$U(N)\sim U(1)\times SU(N)$ symmetry group have been studied in four
dimensions~\cite{creutz1} and the special behavior of the $U(1)$ factor had
been remarked upon--it would be most interesting to study quantitatively the
role of the anisotropy. One expects the $U(1)$ factor to trigger the
appearence of the layered phase~\cite{nicolis07}.

\acknowledgments
It is a pleasure to acknowledge discussions with S.~M.~Catterall,
Ph.~de~Forcrand\linebreak  and  J.~Iliopoulos.

\end{document}